\documentclass[%
 pre, twocolumn,
 amsmath,amssymb,
 aps,
floatfix,
]{revtex4-2}

\usepackage{graphicx}% Include figure files
\usepackage{dcolumn}% Align table columns on decimal point
\usepackage{bm}% bold math
\usepackage{xcolor}
\usepackage{xr}
\usepackage{adjustbox}
\usepackage{hyperref}
\usepackage[mathlines]{lineno}
\hypersetup{
    colorlinks=true,
    linkcolor=blue,
    filecolor=magenta,      
    urlcolor=blue,
    citecolor=blue,
}
\usepackage{color}
\usepackage{listings}
\usepackage{subfigure}
\usepackage{amsthm}
\theoremstyle{definition}

\usepackage{float}  % in preamble
\usepackage{tikz}
\usepackage{amsmath}
\usetikzlibrary{angles, quotes, calc}
\usepackage[normalem]{ulem}

\usepackage{lineno}
\usepackage{amsmath}
\usepackage{mathtools}

\usepackage[makeroom]{cancel}

\usepackage{verbatim}

\begin{document}

\title{Algal Optics}

\author{Ming Yang}
\email[]{my365@cam.ac.uk}
\thanks{Joint first author}
\affiliation{Department of Applied Mathematics and Theoretical 
Physics, Centre for Mathematical Sciences,\\ University of Cambridge, Wilberforce Road, Cambridge CB3 0WA, 
United Kingdom}
\author{Sumit Kumar Birwa}
\email[]{skb61@cam.ac.uk}
\thanks{Joint first author}
\affiliation{Department of Applied Mathematics and Theoretical 
Physics, Centre for Mathematical Sciences,\\ University of Cambridge, Wilberforce Road, Cambridge CB3 0WA, 
United Kingdom}%
\author{Raymond E. Goldstein}
\email[]{R.E.Goldstein@damtp.cam.ac.uk}
\affiliation{Department of Applied Mathematics and Theoretical 
Physics, Centre for Mathematical Sciences,\\ University of Cambridge, Wilberforce Road, Cambridge CB3 0WA, 
United Kingdom}%
\date{\today}% It is always \today, today,

\begin{abstract}
Nearly a decade ago it was discovered that the spherical cell body of 
the alga {\it Chlamydomonas reinhardtii} can act as a lens 
to concentrate incoming light onto the 
cell's membrane-bound photoreceptor and thereby affect phototaxis.  
Since many nearly transparent cells in marine environments have 
complex, often non-axisymmetric shapes, this observation 
raises fundamental, yet little-explored questions in biological 
optics about light refraction by the bodies of microorganisms.  
There are two distinct contexts for such questions: the {\it absorption} 
problem for {\it incoming} light, typified by 
photosynthetic activity taking place in the chloroplasts of
green algae, 
and the {\it emission} problem for {\it outgoing} light, where
the paradigm is bioluminescence emitted from 
scintillons within dinoflagellates.
Here we examine both of these aspects of ``algal optics" in 
the special case where the absorption or emission is localized in 
structures that are small relative to the overall organism size,
taking into account both refraction and reflections at the cell-water boundary.  
Analytical and numerical 
results are developed for the distribution 
of light intensities inside and outside the body, and we establish certain duality relationships
that connect the incoming and outgoing problems.  
For strongly non-spherical shapes we find lensing effects that 
may have implications for photosynthetic activity
and for the angular distribution of light emitted during bioluminescent flashes. 
\end{abstract}
\maketitle

%%%%%%%%%%%%%%%%%%%%%%%%%%%%%%%%%
\section{Introduction}
\label{intro}

\begin{figure}[b]
\includegraphics[width=1.0\columnwidth]{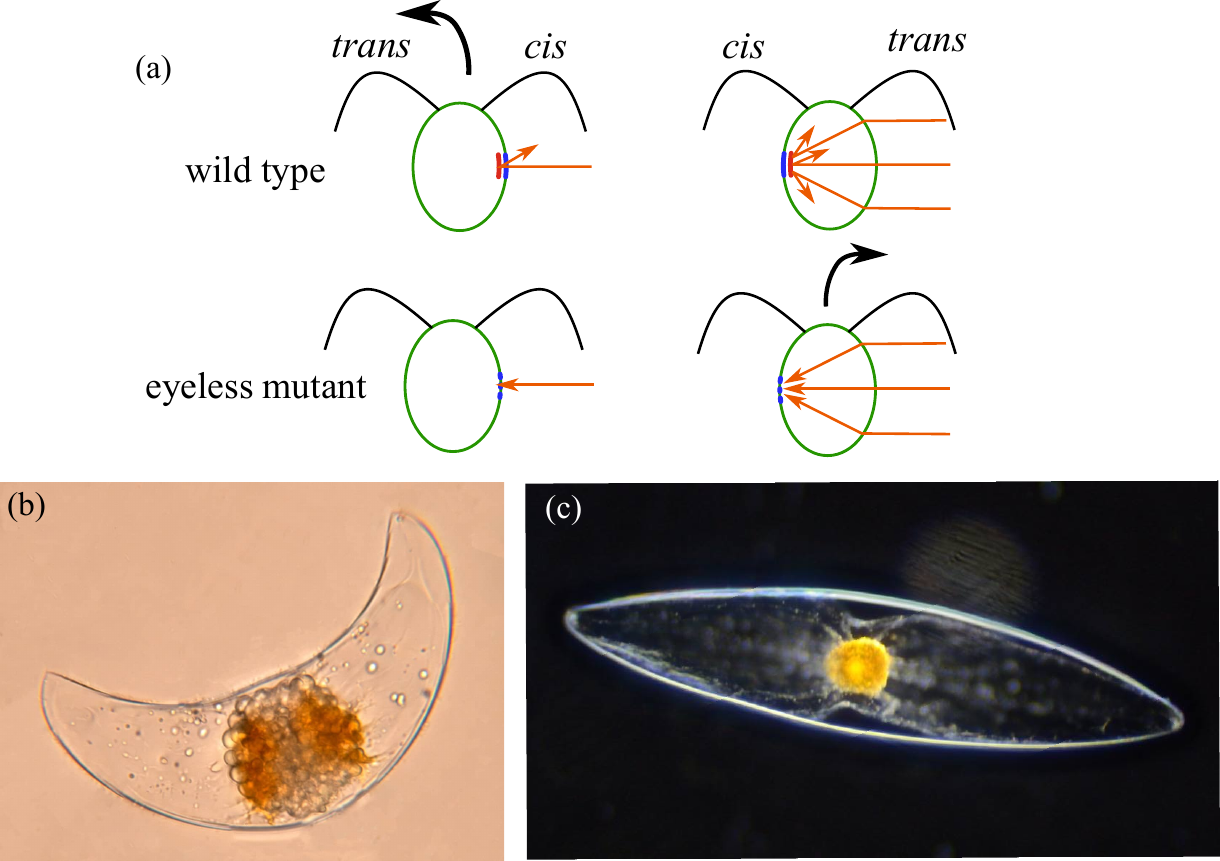}
\caption{Algal optics. 
(a) Photoreceptor shading in wild type and eyeless mutants of \textit{Chlamydomonas reinhardtii}, 
after \cite{Ueki2016}.  Dinoflagellates: (b) crescent-shaped
\textit{Pyrocystis lunula} and (c) spindle-shaped
\textit{Pyrocystis fusiformis} \cite{fusiformis_ref}.}
	\label{fig1}
\end{figure}

In a remarkably prescient paper \cite{Kessler2015}, Kessler, Nedelcu, Solari and Shelton
found in 2015 that the spheroidal cell bodies of various green algae can act as lenses,
bringing light from a distant source to a focus outside the cell.  They
suggested there might be functional significance
to such a lensing effect, and indeed in the following year Ueki, {\it et al.} \cite{Ueki2016}
found the first example, using the green alga {\it Chlamydomonas reinhardtii}.  Phototaxis
in {\it Chlamydomonas} is achieved by the coupling between signals received by 
a photosensor and the beating dynamics of the two flagella that are anchored below the cell
wall near the anterior pole of the cell.   
The 
photosensor (Fig.~\ref{fig1}) sits within a membrane at the periphery of the cell, and in 
wild type cells has behind
it an ``eye spot", a pigmented protein layer 
visible in bright field microscopy.   Acting as a quarter-wave plate \cite{Foster1980}, the eye spot 
reflects incident light and thus blocks light from behind
from the photosensor, such that only light coming from outside the cell is
detected.  It is precisely this directionality that underlies the ability of cells to 
steer toward or away from light \cite{Schaller1997,Leptos2023}.
For example, in negatively phototactic mutants lacking the eye spot, light from behind the cell falls
on the photosensor, and one might naively expect the cell to be unable to perform
phototaxis because of the isotropic detection of light.  Yet, because of the lensing effect of the 
cell the intensity of light falling on the photoreceptor from behind 
is greater than that from the forward direction, giving rise to a distinguishable
signal and thus to net phototactic 
motion (erroneously) toward the light.
It is thus plausible that the eyespot was selected by 
evolution precisely because, in providing directionality to the sensing of light, it 
improves the phototaxis of cells \cite{Jekely_evolution}.  

Lenses and lens-like properties of cells are known in other systems.  For example, 
there is evidence for a lens {\it within} 
the eyespot of the
dinoflagellate \textit{Nematodinium} \cite{Francis} 
that helps to gather light onto its photoreceptor. 
At the same time as the study of Ueki, {\it et al.}, 
separate work on cyanobacteria \cite{Schuergers} 
showed that its cell body can act as a lens, 
creating an uneven distribution of light 
intensity in the inner cell wall. 
Further work showed that a type of supermolecular
machine for cell mobility is activated by that 
directional light \cite{Nakane}.  These kinds of 
effects are now broadly understood for 
prokaryotes \cite{Wilde}.
Finally, we note the case of diatoms, algae whose 
surface has a complex silica microstructure 
patterned on the scale of the wavelength of 
visible light. These structures induce
wavelength-dependent optical properties \cite{Ghobara, Tommasi} and diffraction \cite{Maibohm}.

Taking a broader view, there are important historical 
examples where lensing effects are thought to impact on 
life's processes, particularly with regard to plants \cite{Vogelmann}.  
For example, phototropism, the movement of plants toward light,
is dependent on the fact that the cytoplasm of plant cells has a
higher index of refraction than the surrounding medium 
(air or water).  This was demonstrated by an experiment 
\cite{Humphry}
that showed that this movement reverses direction when 
a plant is immersed in a medium with higher index of 
refraction than the cytoplasm.  
Many investigations have focused on the effect of cell shape on 
photosynthesis in plants and fungi. For example, a study \cite{Dennison} of the
filamentous fungus \textit{Phycomyces blakesleeanus} showed
that the light intensity at the cell 
surface is enhanced by a factor of $\sim 2$, 
while theoretical work suggests an even larger 
boost \cite{Myneni}.  The epidermal cells of 
certain tropical plant species are thought to act as lenses, providing an advantage in 
light-gathering ability for shade plants \cite{Bone}, although later work found that 
the structure does not help to gather diffuse 
light \cite{Brodersen}.

The results summarized above indicate that lensing effects by the
cell bodies of microorganisms can have a functional significance, even in the simplest 
possible geometry of a sphere.  
Yet, there are many freshwater and marine microorganisms with strikingly nonspheroidal shapes, as
exemplified by dinoflagellates such as {\it Pyrocystis lunula} and {\it Pyrocystis fusiformis} 
shown in Fig.~\ref{fig1}(b,c).  The properties of such complex, often non-axisymmetric 
bodies appear to be completely unknown in the biological optics literature, 
although there is now a growing field of 
``freeform optics" \cite{freeform1,freeform2,freeform3,freeform4} that considers nonaxisymmetric shapes.  
Motivated by these strange
and wondrous forms, in this paper we commence an in-depth study of the field of ``algal optics".

Beyond their phototaxis, organisms such as green algae and dinoflagellates are 
photosynthetic and their chloroplasts, the organelles containing the photosynthetic apparatus, 
are in characteristic positions within the cell.  In the case of dinoflagellates, the chloroplasts move 
around the cell in diurnal
patterns \cite{SwiftTaylor,Sweeney,Heimann}, changing the absorption profile of the 
cytoplasm \cite{Stephens}.  A natural question
is whether lensing can enhance the 
intensity of light falling on chloroplasts.  This is the ``incoming" problem.

Dinoflagellates are among the many
marine and freshwater organisms that
exhibit bioluminescence  
\cite{Haddock}.  Unlike the steady glow of bioluminescent
bacteria, these eukaryotes emit bright flashes of light
in response to fluid or mechanical shear \cite{Latz,Jalaal}.
This light emanates from membrane-enclosed organelles
termed ``scintillons", within which occur chemical 
reactions involving the protein luciferin.  A second natural
question is thus whether lensing can alter the
spatial distribution of light emitted from such sources.  This is the ``outgoing" problem.

\begin{figure}[t]
\includegraphics[width=0.98\columnwidth]{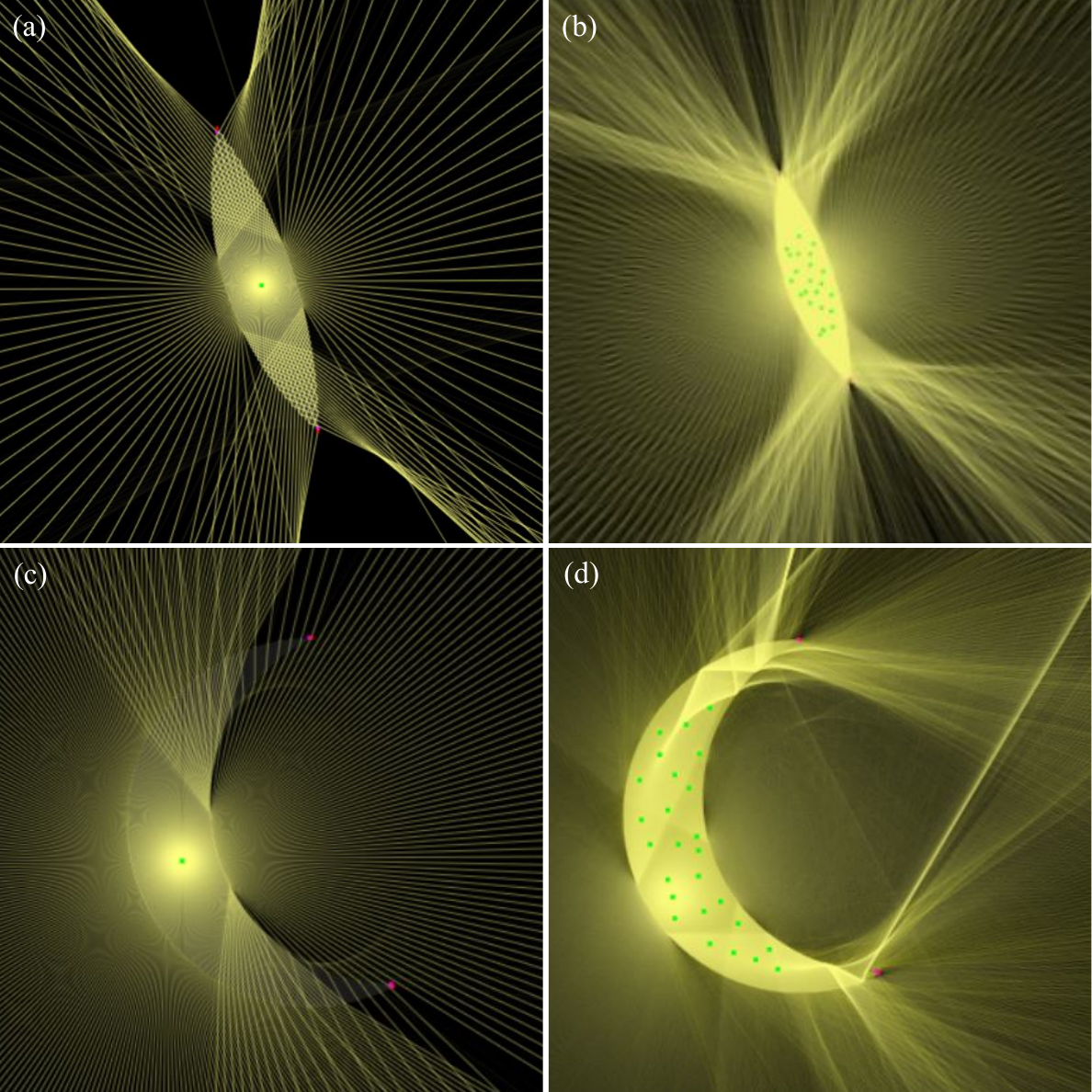}
\caption{Ray tracings for various dinoflagellate geometries \cite{rayoptics}. Single and 
multiple sources for the 
shape of \textit{P. fusiformis} (a,b) and of \textit{P. lunula} (c,d).}
\label{fig2}
\end{figure}

Figure~\ref{fig2} presents ray-tracings 
using a ray optics simulator \cite{rayoptics} 
that illustrate how light is emitted from internal
sources in different dinoflagellate cell shapes. 
The geometries represent the spindle-shaped cell 
body of \textit{P. fusiformis} (a,b) and the crescent-shape of \textit{P. lunula} (c,d). 
The green dots indicate the positions of bioluminescent sources within the cell body.
In (a,b) the emitted rays are symmetrically 
distributed about the long axis, with intensity 
primarily directed laterally\textemdash 
toward the left and right\textemdash and little 
light escaping along the axial (top–bottom) 
directions. This symmetry and directional focus 
suggest that the fusiform shape acts as a 
cylindrical lens, favoring horizontal emission. 

In the single-source case for the crescent shape 
(c), the emission exhibits a markedly different pattern; light is preferentially emitted along 
the axial directions (top and bottom), forming 
focused streaks of high intensity. Moreover, 
there is an enhanced concentration of rays on 
the concave side of the cell, particularly close 
to the cell body, due to the curvature-induced 
refraction. In the multiple-source case (d), the 
overall emission appears more spatially diffuse, 
but the asymmetry persists, 
with noticeably more rays directed toward the 
concave side. The convex side emits less light 
overall, likely due to the outward bend 
redirecting rays away from those regions.
These results highlight the critical role of geometry in shaping the angular emission profile. 
Even in the absence of specialized optics, the 
cell’s shape can redistribute internally generated 
light in a highly non-uniform manner. 

\begin{figure}[t]
\includegraphics[width=0.98\columnwidth]{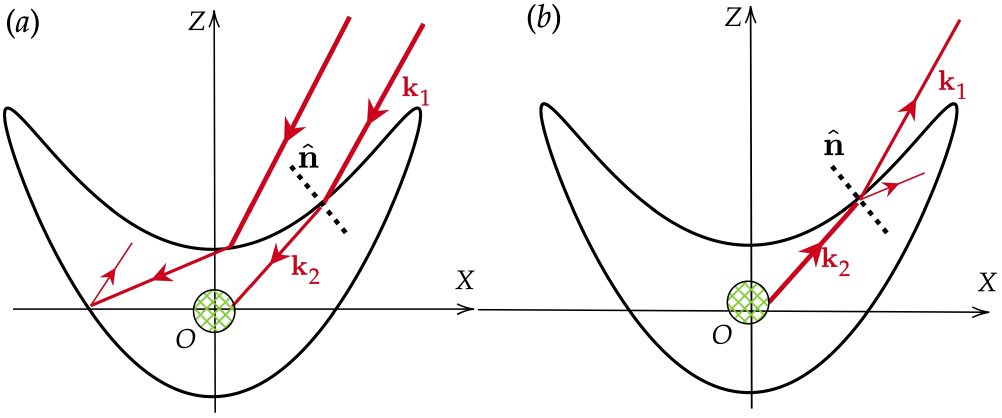}
\caption{The two optical problems discussed here. Red lines are principal light rays. 
(a) The incoming problem of how cell shape influences intensity
of photosynthetic light falling on a chloroplast. (b) The outgoing problem explores 
how the bioluminescence emission is influenced by 
the cell geometry.}
\label{fig3}
\end{figure}

In this paper we study the incoming and outgoing problems with analytical and numerical methods. 
For the motivational cases of green algae and dinoflagellates there is a reasonable separation between
the size of the absorber or emitter (the algal photoreceptor is $\sim 1-2\,\mu$m across, 
scintillons are $\sim 0.5-1.5\,\mu$m in diameter) and the size of the entire cell
(\textit{Chlamydomonas} is $\sim 10\,\mu$m cross, dinoflagellates can be $\sim 150\,\mu$m long).
Given this, in the simplest model we consider absorbers/emitters to have a radius $a\ll R$, where
$R$ is a characteristic size of the cell.  
Section \ref{sec:preliminaries} sets up the incoming and 
outgoing problems in mathematical terms, laying out the additional modeling assumptions and 
definitions.  As an example of the problem of interest, the observations of Ueki, et al. are 
analyzed quantitatively to gain insight into the lensing effects that can occur. 
Analytical results for two-dimensional bodies are presented in \S~\ref{sec:2Danalytical}, including 
the results of averaging over orientations. 
Three-dimensional problems are considered analytically 
in \S~\ref{sec:3Danalytical} and numerically in \S \ref{sec:3Dnumerical}, where we 
illustrate strong, complex lensing
effects associated with shapes like those of the genus \textit{Pyrocystis}.  In \S \ref{sec:discussion}
we discuss possible experiments to examine this problem in greater detail.

\section{Preliminaries}
\label{sec:preliminaries}

We consider two complementary problems, each motivated by considerations of the natural environment.

\textit{The incoming problem: How cell geometry affects light absorption in photosynthesis.}  
As shown in Figure~\ref{fig3}(a), this problem explores how a curved cell wall modifies the 
light intensity at a specific 
location inside the cell. In the turbulent ocean, microorganisms receive sunlight that is 
scattered and refracted from nearly 
all directions, while their random orientations within the flow further homogenize the 
light distribution.   We consider the case in which the 
cell geometry and random flows are such that there is a uniform angular distribution of 
cell orientations, so the incoming 
light appears isotropic from the cell's perspective.  The mathematical problem of interest 
is then the light intensity received by a small 
target within the cell relative to that in the absence of the surrounding cell.

\textit{The outgoing problem: How cell geometry shapes bioluminescence emission.}  
In marine environments, suspended dinoflagellates respond to the fluid flows 
associated with ambient turbulence and disturbances from large 
predators by giving off bioluminescent flashes.  Figure~\ref{fig3}(b) illustrates 
the mathematical problem of interest: 
how a cell's shape alters the angular distribution of bioluminescent light emitted 
isotropically from small sources within the cell, 
undergoing both refraction and internal reflection. Since the cell 
is much smaller than the distance to predators, the focus is on the direction of 
emitted light rather than its origin. 

\begin{table}[]
    \begin{tabular}{|c|c|}
        \hline
        Definition & Variable \\ \hline
        Incident/outgoing light rays & $\textbf{k}_1$, $\textbf{k}_2$ \\ \hline
        Incident angle & $\theta_i$ \\ \hline
        Transmitted angle & $\theta_t$ \\ \hline
        Reflected angle & $\theta_r=\theta_i$ \\ \hline
        Energy transmission, Eq.~\eqref{eq:fresnel} & $f$ \\ \hline
        Intensity boost & $\eta$\\ \hline
        Local surface normal & $\textbf n$ \\ \hline
        Local radius of curvature &$R$\\ \hline
        Relative refractive index , Eq.~\eqref{def:snell}&$n$\\ \hline    
    \end{tabular}
    \caption{List of variables and their definitions.}
    \label{tab:my_label}
\end{table}

Our analysis assumes that the interior of a photosynthetic cell is a homogeneous 
optical medium with refractive index $ n_{\text{cell}} $, while the surrounding aqueous environment 
has refractive index $ n_{\text{water}} \simeq 1.33 $. For example, \textit{Chlamydomonas} cells 
have been reported to exhibit $ n_{\text{cell}} \simeq 1.47 $ in the visible spectrum~\cite{Ueki2016}, 
which implies a relative refractive index $n = n_{\text{cell}}/n_{\text{water}} \simeq 1.1$. We adopt 
this value in our calculations as representative of green algal exposed to light in aqueous environments.

To model how sunlight enters a cell and contributes to photosynthesis, we apply
Snell's law to incident light rays arriving from the surrounding water
\begin{equation}\label{def:snell}
    \sin \theta_i = n \sin \theta_t,
\end{equation}
where $\theta_i$ is the angle of incidence in water, and $\theta_t$ is the 
angle of transmission inside the cell, both measured with respect to the local normal.

The fraction of energy transmitted into the cell 
quantifies how much unpolarized light passes 
through the interface. For incident angles 
$\theta_i < \theta_c = \sin^{-1}(1/n)$, where 
total internal reflection does not occur, the 
Fresnel transmission coefficient $f(\theta_i)$
for unpolarized light is \cite{Jackson}
\begin{equation}\label{eq:fresnel}
    f(\theta_i) = 1 
    - \frac{1}{2} \left( \frac{\sin^2(\theta_i - \theta_t)}{\sin^2(\theta_i + \theta_t)} + 
    \frac{\tan^2(\theta_i - \theta_t)}{\tan^2(\theta_i + \theta_t)} \right),
\end{equation}
where $\theta_t = 
\arcsin\left(\sin \theta_i/n\right)$ is the transmitted angle determined by Snell's law.
Normal incidence ($\theta_i = 0$) gives the maximum transmittance rate $f(0) = 4n/(n + 1)^2$. 
The quantify $1 - f(\theta_i)$ is the proportion 
of light reflected back into the environment; it determines how
much light penetrates the cell for internal processes.

\begin{figure}[t]
    \centering
    \includegraphics[width=\columnwidth]{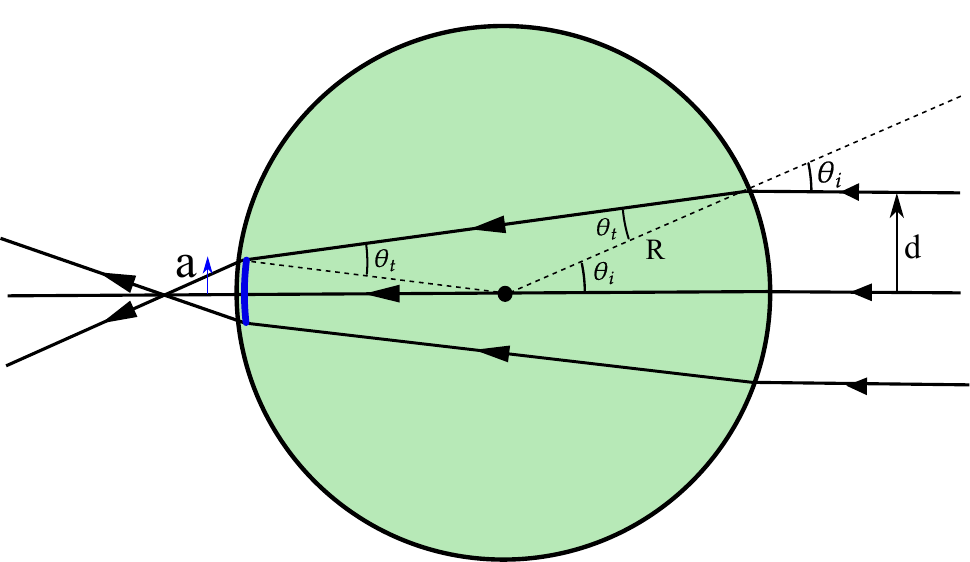}
    \caption{Analysis of the eyespot mutant in \textit{Chlamydomonas}. 
    The photosensor (blue) is located at the inner boundary of the cell. Although incoming light rays focus 
    outside the cell, a gathering effect is experienced by the photosensor.}
    \label{fig4}
\end{figure}

We define the \textit{boost factor} $\eta$ as a measure of light amplification induced by the cell’s 
geometry. For the incoming problem, $\eta = d\Omega_2 / d\Omega_1$ is the ratio of differential solid angles 
$d\Omega_i$ subtended at the entry and focus points, capturing how ray convergence enhances 
local light intensity. 
For the outgoing problem, $\eta = dA_1 / dA_2$ expresses the relative change in projected beam 
area as light exits the cell, reflecting how geometry reshapes outgoing flux.

To simplify the analysis of light-ray interactions, we adopt the \textit{chief-ray approximation}, 
in which all rays are assumed to deviate only slightly from a central principal ray. This reduces 
geometric complexity while preserving essential directional and focusing behavior.

\subsection*{Example: Optical Boost at the Algal Eyespot}

We begin with a quantitative analysis of the 
results of Ueki, et al. \cite{Ueki2016} on 
the \textit{eyeless} mutant of 
\textit{C. reinhardtii}. 
Fig.~\ref{fig4} shows the geometry;
a cell of radius $R$ has a photosensor 
of radius $a$ at its
periphery, modeled as a circular patch.
Light enters the 
cell from the opposite side, refracts at the 
surface, and the rays converge past the cell. 
A cone of these rays hits the photoreceptor. 
The maximum angular deviation of rays that 
reaches the photoreceptor defines a limiting 
incident angle $\theta_i$ with ``impact 
parameter" $d=R\sin\theta_i$ as in classical scattering theory.

Figure~\ref{fig4} shows the geometry of interest. 
The vertex angle of the large isosceles triangle is
$\pi-2\theta_t$, so if we add all the vertex angles of 
triangles touching the center we find
$a/R+\pi-2\theta_t+\theta_i=\pi$, and 
obtain a relation 
between the ratio $\epsilon\equiv a/R$ and 
the incident and refracted angles,
\begin{equation}
    \epsilon= 2\theta_t-\theta_i\simeq
    \frac{2-n}{n}\theta_i.
\end{equation}
where the second relation follows from Snell's law 
and is valid for 
$\epsilon \ll 1$, where a  
small-angle approximation holds.

The optical boost $\eta$ 
is the ratio of areas of the incident light
cone to that of the photoreceptor patch.
In two dimensions, and in the small-angle 
approximation, this corresponds to the 
length ratio
\begin{equation}
    \eta_{\mathrm{2D}} = \frac{2d}{2a} 
    \simeq \frac{\theta_i}{\epsilon}
    =\frac{n}{2-n}.
    \label{2Deta}
\end{equation}
    
In three dimensions, the incoming rays span a 
circular disk of radius $d$, and the receiving
region is a disk of radius $a$, so the boost becomes the area ratio
\begin{equation}
    \eta_{\mathrm{3D}} = \frac{\pi d^2}{\pi a^2} 
    = \eta^2_{\mathrm{2D}}
    =\left(\frac{n}{2-n}\right)^2
    \approx 1.49,
    \label{3Dboost}
\end{equation}
where we have used the value 
$n = 1.1$ for \textit{Chlamydomonas}.
This remarkably simple result shows 
that the index of refraction 
alone determines the boost 
for this simple spherical geometry.  The quantity $\eta-1$,
the additional flux of light onto the photoreceptor, is
\begin{equation}
    \eta_{\mathrm{3D}}-1 
    \simeq \frac{4(n-1)}{(2-n)^2},
\end{equation}
which is positive only
when the relative index $n>1$.

The large boost in Eq.~\eqref{3Dboost} implies that an eyeless \textit{Chlamydomonas} cell 
experiences two very distinct signals during each rotation about its body-fixed axis, the one from 
behind being $50$\% stronger than the other.  Not surprisingly, that significantly larger signal 
dominates and the cell moves opposite to the wild type.  The phototactic dynamics of eyeless mutants 
are discussed elsewhere \cite{eyeless}.

\section{Analytical Results for 2D Bodies}
\label{sec:2Danalytical}

In this section we study the incoming 
problem for 2D bodies, 
progressing from simple to complex.
We scale lengths by the size
$R$ of the body, so that the small target has radius $\epsilon=a/R$ and the bundle of
light rays that intersects the target
has half width $\delta=d/R$.

\begin{figure}[t]
\includegraphics[width=0.95\columnwidth]{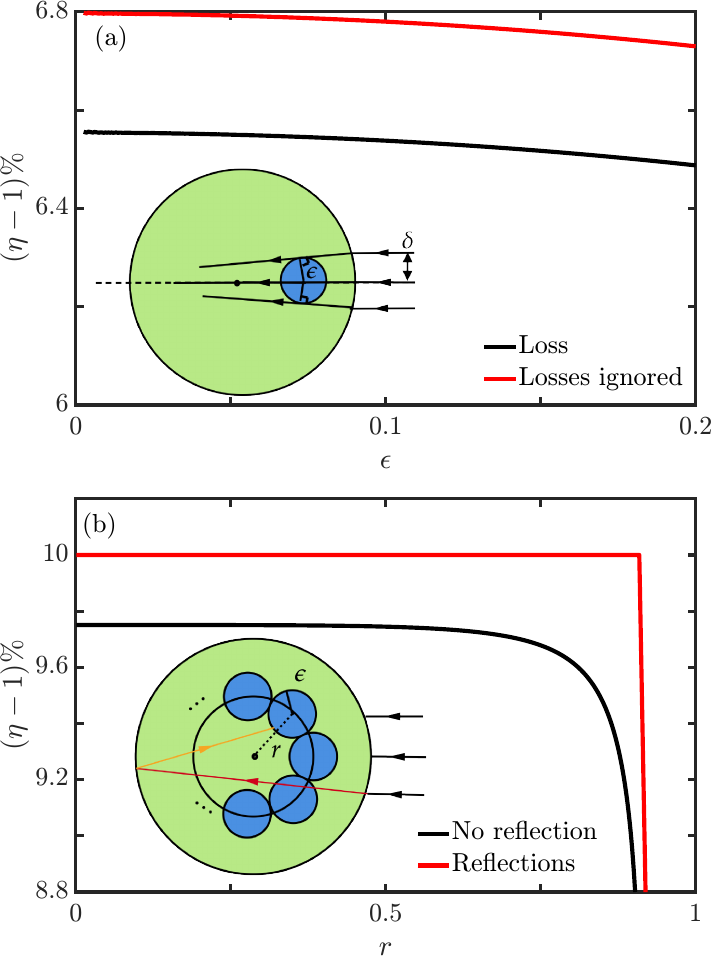}
\caption{Incoming problem for a circle. 
(a) 
Boost versus ball size for $r = 0.3$ and relative index $n=1.1$. 
As $\epsilon$ increases, $\eta$ decreases, 
indicating weaker lensing and greater losses \eqref{eq:fresnel} when incident angle is large. 
(b) Disks of radius $\epsilon$ densely populate a 
circular shell. When multiple internal reflections 
are included, a ray (red) reaching the second disk 
is counted via the shell’s 
absorption. 
The boost is constant when all 
reflections (orange line) are included, until the ball reaches the total internal 
reflection zone where increasing $r$ can no longer gain more energy. If reflections 
are excluded, $\eta$ drops more near the upper
boundary  due to increased losses.} 
\label{fig5}
\end{figure}

\textbf{A. The Circle.} We begin with the simplest case, 
a homogeneous circular body, which serves as a baseline for 
understanding how light refracts and concentrates within
2D bodies. Figure \ref{fig5}(a) 
shows the setup: a small, fully absorbing 
test ball of radius $\epsilon$ is at position 
$(r, \varphi)$ where $r\in[0,1]$ is the (scaled) radial distance 
from the center and $\varphi$ is the angle relative 
to the incoming light direction. 

We first consider the 
case where the test ball is 
at the center of the circle to evaluate 
the maximal focusing effect.
Because of the circular symmetry, we may take 
the incoming rays to be parallel and incident from the  
$+x$ direction without loss of generality.  
The boundary entry points of rays that ultimately 
graze the edges of the test ball define the incident 
angular window. If a grazing 
ray enters at a point on the circle at angle $\theta_i$
then, since the normal 
to the circle at that point is 
radial, we have $\delta=\sin\theta_i$. And as 
the interior angle between the refracted ray 
grazing the target and the radial line from the origin 
that intersects the points of entry is $\theta_t$ we 
have the second relation
$\epsilon = \sin\theta_t$,
and thus $\delta = \epsilon n$.
Thus, the effective ``window” of rays that can 
reach the test ball spans a lateral width $\delta$ 
that increases with the relative refractive index $n$. Note that the distance $\delta\leq1$, 
so the corresponding boost $\eta$, 
the ratio of this width to the width of the test ball, is
\begin{equation}\label{eq:eta}
\eta = \frac{\delta}{\epsilon} = \min\{n,1/\epsilon\}.
\end{equation}

The boost equals the relative refractive index 
until it reduces because the ball can at most 
receive light with half-width $\delta=1$. The 
transition is also when total internal reflection 
occurs. This clean result captures the idealized 
case with 
perfect transmission and no optical loss, showing that 
the absorption at the center scales linearly with the 
refractive index $n$. In reality, partial reflection 
occurs at the boundary, especially for rays striking 
at oblique angles, due to refractive mismatch with the surrounding medium.

To incorporate such losses, we incorporate the 
angle-dependent transmission coefficient 
$f(\theta_i)$ 
approximated using 
Eq.~\eqref{eq:fresnel} for unpolarized light. 
The net boost factor, corrected for reflection losses, is then
\begin{equation}\label{eq:fresnel:eta}
\eta = \frac{1}{2\epsilon}\int_{-\theta_{\max}}^{\theta_{\text{max}}}\!\! d\delta(\theta_i)\, 
f(\theta_i)=\frac{1}{\epsilon}\int_0^{\theta_{\max}} \!\! d\theta_i\, f(\theta_i)\cos \theta_i,
\end{equation}
where $\theta_{\max}=\max\{\sin^{-1}(n \epsilon),\pi/2\}$ is the 
angular window of rays that refract to 
the center and $\cos\theta_i$ is the Jacobian $\partial\delta(\theta_i)/\partial \theta_i$. 
Eq.~\eqref{eq:fresnel:eta} reduces to \eqref{eq:eta}
when $f=1$. 

\textbf{B. Effect of Absorber Size.} We now consider 
test ball locations with $\varphi=0$, so the geometry 
remains symmetric about the $x$-axis, but with $r>0$. 
The boundary entry points of rays that graze the ball 
define a narrow angular range of incident rays. Using 
the geometry in Fig.~\ref{fig5}(a), 
the condition for a ray to reach the disk is 
\begin{equation}
r\sin(\theta_i - \theta_t) + \sin\theta_t=\epsilon,
\label{rtheta}
\end{equation}
which implicitly defines $\theta_i$ as a 
function of $\epsilon$ and $r$.
Once $\theta_i(\epsilon)$ is known, the boost factor 
$\eta$ is calculated from Eq.~\eqref{eq:fresnel:eta}. The resulting boost profile is plotted in 
Fig.~\ref{fig5}(a) for $r=0.3$. The boost is always smaller than the $10\%$ boost 
when the ball is placed at the center, as in Eq.~\eqref{eq:eta}, because lensing is 
weaker closer to the light source.
In the small-angle limit, \eqref{rtheta} yields the boost
\begin{equation}
    \eta\simeq \frac{n}{1+r(n-1)},
\end{equation}
which interpolates between the limit $n/(2-n)$ in \eqref{2Deta} for $r=-1$, $n$ for $r=0$ and $1$ 
as $r\to 1$.
The boost profile is continuous and converges to a constant value as $\epsilon \to 0$. This validates the 
use of small but finite test balls in simulations to estimate local light intensity. In biological terms, 
this corresponds to evaluating the light absorption by a small chloroplast placed at a 
given location within the cell. 
Because diffraction and coherence effects are negligible at this scale, the ray-based results 
are expected to 
agree with those from wave-optics models in the geometric optics limit.

\textbf{C. Angular Averaging.} In natural 
biological contexts, light rays are likely 
to come from all direction relative to a cell.  The results above
should then be averaged of the distribution of incoming light rays, 
and the simplest assumption is a uniform distribution 
of the angle $\varphi\in [0, 2\pi]$. 
We can compute the boost at 
each angle, and perform this average, or equivalently,
we can densely distribute $N$ replicas of the test ball 
within a circular shell at radius $r$, with each 
ball interacting with light rays from different 
directions as illustrated in Fig.~\ref{fig5}(b).  

The circumference of the shell is the sum of the diameters of all the balls, so we have 
$2 \epsilon N = 2 \pi r$. The total energy $E_{\text{tot}}$ absorbed by the 
shell and the average energy $E_{\text{avg}}$ 
absorbed by an individual test ball are related by
\begin{equation}\label{eq:tot vs avg}
E_{\text{tot}} = \alpha N E_{\text{avg}},
\end{equation}
where $\alpha$ is inverse of the averaged number 
of times a light ray intersects a ball before reaching 
the cell wall. We will show that the factor 
$\alpha=1/\pi$ in two ways. First, note that the 
average energy absorbed is proportional to the 
circumference, as proved below in Eq.~\eqref{eq:averaged:projection:2D}, 
so $E_{tot}/N E_{avg}=2\pi r/2\pi \epsilon N$, indicating the result. Second, 
if we take the limit $r \to 0$ while maintaining 
$\epsilon/r$ fixed to be a small number we are implementing 
a similarity transformation in which all ratios 
remain the same, and Eq.~\eqref{eq:tot vs avg} still 
holds, with the same $\alpha$. In this limit, all small balls approach the center, 
at which the total energy is the linear function $E(r)=2 rn f(0)$, so
\begin{equation}\label{eq:19}
\frac{E_{\text{avg}}}{E_{\text{tot}}} = \frac{\epsilon}{r},
\end{equation}
which then yields $\alpha=1/\pi$. 

The light intensity profile as a function of the distance from the center of the circle 
is shown in Fig.~\ref{fig5}(b), using  
the energy-averaging framework above. At each radial position $r$, the shell can be treated as a disk 
of radius $r$ centered within the cell, and the local boost $\eta(r)$ is computed using 
Eq.~\eqref{eq:fresnel:eta}, with $\theta_{\text{max}}=\max\{\sin^{-1}(nr),\pi/2\}$. As $r \to 1$, the shell approaches 
the boundary of the circle, and the transmitted light at large $\theta_i$ experiences greater loss when 
crossing the cell wall, as described by Eq.~\eqref{eq:fresnel}. This results in a decrease in $\eta(r)$, 
illustrated by the black curve in Fig.~\ref{fig5}(b), which drops sharply beyond $r\approx 0.75$.

We extend this energy absorption analysis to include an arbitrary number of internal reflections for the test ball. 
As before, the shell around the central disk is densely populated with smaller disks. These 
do not absorb light rays directly, as the presence of one does not interfere with another disk intercepting 
the same ray. Thus, we simply track the number of times a light ray passes through the shell. 
The total absorbed energy consists of the initial contribution from the ray passing through the cell wall, 
along with all subsequent internal reflections (see illustration in Fig.~\ref{fig5}(b), 
where the thin red ray represents the first reflection). By symmetry, each reflected ray follows 
the same path as the initial one, preserving its angle of incidence.

Denoting the incoming light ray energy at angle 
$\theta_i$ as $E(\theta_i)$, the total energy 
absorbed by the shell is
\begin{align}
E_{\text{tot}}(\theta_i) &= E(\theta_i) f(\theta_i) + 
E(\theta_i)f(\theta_i)(1 - f(\theta_i)) + \nonumber \\ 
& \quad \quad \quad \quad E(\theta_i) f(\theta_i)(1 - f(\theta_i))^2 + \dots \nonumber\\
&= \frac{E(\theta_i) f(\theta_i)}{1 - (1 - f(\theta_i))} = E(\theta_i).
\end{align}
Thus, the contributions from all internal reflections exactly cancel the transmission losses, 
resulting in a constant boost given by Eq.~\eqref{eq:eta}, with $\eta = \max\{n,1/r\}$. This is illustrated 
by the red curve remaining flat until it drops significantly beyond $r=1/1.1$ in Fig.~\ref{fig5}(b).

\textbf{D. Duality for a Circle.} 
In 2D, it can be shown explicitly that the incoming and outgoing problems are equivalent. 
We first prove 
this result for a circle without  losses~\eqref{eq:fresnel}, and then proceed  to the general case.

\begin{figure}[t]
     \includegraphics[width=1\linewidth]{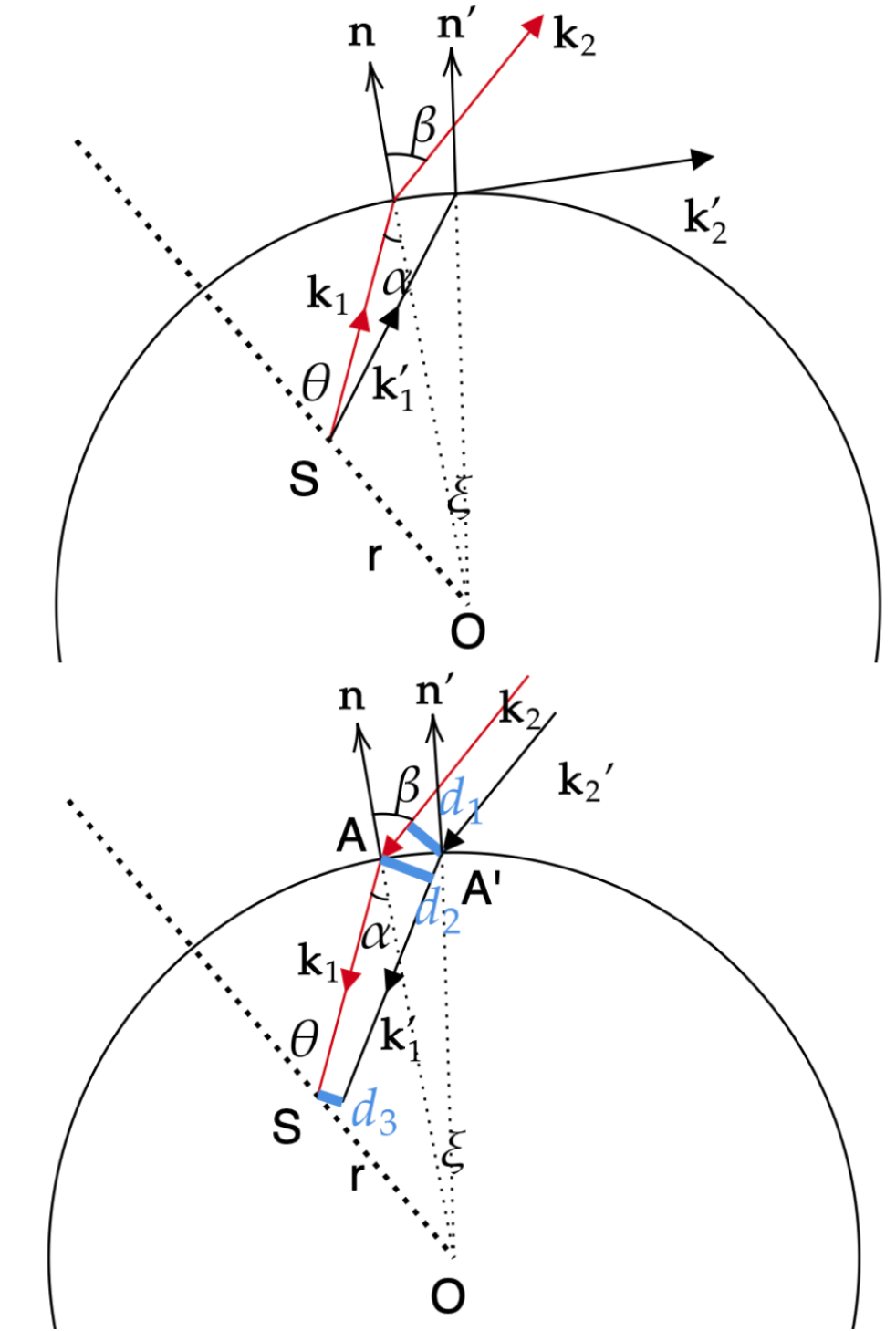}
    \caption{Geometry to demonstrate duality. (a) Bioluminescence case. 
    Principal light rays are shown in red. We omit the angles between the light rays shown in 
    black, denoted by $\theta', \alpha'$ and $ \beta'$, (b) Photosynthesis case. Incoming 
    rays span a length $d_1$, and the absorber has a projected length $d_3$.}
    \label{fig6}
\end{figure}

Consider first the outgoing problem.  We adopt the notation 
$\angle(\mathbf{u}, \mathbf{v})$ for the
angle between vectors $\mathbf{u}$ and 
$\mathbf{v}$.
With reference to Fig. \ref{fig6}(a), the geometric boost for bioluminescence 
$\eta_B$ is defined as the limiting 
ratio of angles
\begin{equation}\label{eq:app:13}
    \eta_B(\theta) = \lim_{\angle(\mathbf{k}_1, \mathbf{k}_1') \to 0} 
    \frac{\angle(\mathbf{k}_1, \mathbf{k}_1')}{\angle(\mathbf{k}_2, \mathbf{k}_2')}.
\end{equation}
With $\alpha=\angle(\mathbf{k}_1, -\mathbf{n})$, $\beta=\angle(\mathbf{k}_2, \mathbf{n})$, and 
$\theta=\angle(\mathbf{k}_1, \textbf{SO})$, 
trigonometry and Snell's Law~\eqref{def:snell} yield
\begin{equation}\label{eq:app:14}
    \frac{\sin\alpha}{\sin\theta}=r \quad {\rm and} \quad \frac{\sin\alpha}{\sin\beta}=\frac{1}{n}.
\end{equation}
For the primed vectors and angles we have
the analogous relations, 
\begin{equation}
    \frac{\sin\theta'}{\sin\beta'}=\frac{1}{rn}
    \quad {\rm and} \quad \frac{\sin\theta'}{\sin\alpha'}=\frac{1}{r},
\end{equation}
leading to relations for 
small changes in the angles,
\begin{align}\label{eq:app:16}
     \beta'-\beta &\simeq  nr\frac{\cos\theta}{\cos\beta}(\theta'-\theta)  \\
     \alpha'-\alpha &\simeq r\frac{\cos\theta}{\cos\alpha}(\theta'-\theta).
\end{align}
Let $\gamma = \angle(\mathbf{k}_2, {\bf OS})=\theta-\alpha+\beta$ and similarly define 
$\gamma'$. Then from geometry we find
\begin{equation}
    \angle(\mathbf{k}_2, \mathbf{k}_2') = \gamma'-\gamma=\theta'-\theta+\alpha-\alpha'+\beta'-\beta,
\end{equation}
and hence, using Eq.~\eqref{eq:app:16} we 
obtain the boost~\eqref{eq:app:13}
\begin{align}
    \eta_B(\theta) &= \frac{\theta'-\theta}{\theta'-\theta+\alpha'-\alpha+\beta'-\beta}\nonumber\\
    &=\frac{1}{n}\left(\frac{\sin\alpha\sin(\theta-\alpha)}{\sin\beta\sin\theta\cos\alpha}+\frac{\cos\theta\sin\alpha}{\cos\beta\sin\theta}\right)^{-1},
\end{align}
where in the last line we have used Eq.~\eqref{eq:app:14} and a product-to-sum trigonometric identity.

For the incoming (photosynthetic) case 
shown in Fig.~\ref{fig6}(b), let point A denote 
the intersection of the refracted ray $\mathbf{k}_2$ with the cell wall, and similarly for A'. The 
boost is the ratio of projected areas
\begin{equation}\label{eq:app:19}
    \eta_P(\theta) = \lim_{d_1 \to 0} \frac{d_1}{d_3}.
\end{equation}
where $d_1$ is the distance between the light rays $\mathbf{k}_2$ and $\mathbf{k}_2'$, 
$d_2$ is the distance between point A and $\mathbf{k}_1'$. $S'$ is a point in the light ray 
$\mathbf{k}_1'$ with ${\bf SS'}\perp {\bf SA}$, and 
$d_3=|{\bf SS'}|$. Also let $d_3'$ be the distance between S and $\mathbf{k}_1'$. 

Since $\xi=\angle({\bf OA}, {\bf OA'})\ll 1$, we have 
$\angle(\mathbf{n},{\bf AA'} )\simeq \pi/2$ and
$|{\bf AA'}|\simeq\xi$.
Geometry then imposes the relations 
\begin{equation}\label{eq:app:20}
    d_1 = \xi\cos\beta, \quad d_2 = \xi\cos\alpha,
\end{equation}
and 
\begin{equation}\label{eq:app:21}
    d_3\simeq d_3'=\xi\cos\alpha-|{\bf SA}|\angle(\mathbf{k_1},\mathbf{k}_1').
\end{equation}
To work out the angle $\angle(\mathbf{k_1},\mathbf{k}_1')=\xi-(\alpha-\alpha')$,
we apply Snell's Law~\eqref{def:snell}
\begin{equation}
    \sin\beta=n\sin\alpha, \quad\sin\beta'=n\sin\alpha'.
\end{equation}
Similar to Eq.~\eqref{eq:app:16}, we obtain
\begin{equation}
    \alpha-\alpha'\simeq \frac{\cos\beta}{n\cos\alpha}(\beta-\beta').
\end{equation}
Combining with $\xi=\beta-\beta'$, we find
\begin{equation}
    \angle(\mathbf{k_1},\mathbf{k}_1')=
    \xi\left(1-\frac{\cos\beta}{n\cos\alpha}\right),
\end{equation}
from which we obtain, using Eq.~\eqref{eq:app:20} and~\eqref{eq:app:21} in~\eqref{eq:app:19},
\begin{equation}\label{eq:app:A:6}
    \eta_P(\theta) 
    =\left(\frac{\cos\alpha}{\cos\beta}-\frac{\sin(\theta-\alpha)\sin(\beta-\alpha)}
    {\sin\theta\sin\beta\cos\alpha\cos\beta}\right)^{-1}.
\end{equation}
Trigonometry identities then lead to the duality result 
\begin{equation}\label{eq:duality2d}
    \eta_P(\theta) = n \eta_B(\theta),
\end{equation}
namely, the intensity profiles of the two cases 
are identical up to a factor of $n$. This is the two-dimensional analogue of ``{\'e}tendue" 
discussed below in \S~\ref{sec:3Danalytical}. 

The outgoing light ray traces the full $2\pi$ angle:
\begin{equation}\label{eq:A9}
    \int_0^{2\pi} \frac{1}{\eta_B(\theta)} d\theta = 2\pi.
\end{equation}
Then our definition~\eqref{eq:app:13} is
$d\theta/d\gamma = \eta_B(\theta)$,
leading to conservation of averaged photosynthesis boost,
\begin{align}\label{eq:A11}
    \eta_P^{\text{avg}}&=\frac{1}{2\pi} \int_0^{2\pi} \eta_P(\theta(\gamma)) d\gamma \nonumber\\
    &=\frac{1}{2\pi} \int_0^{2\pi} \eta_P(\theta(\gamma)) \frac{d\theta}{\eta_B(\theta)} =n.
\end{align}
Here the light rays come at an angle $\gamma$ with ${\bf OS}$, uniformly distributed in $[0,2\pi]$.

\textbf{E. Duality for an Arbitrary 2D Body.}  
For general shapes, any local region can be approximated as an arc segment with a specific radius of 
curvature. The argument applies universally for $r > 1$ (when the point lies outside the circle) or 
$r < 0$ (when the shape is concave). Consequently, the incoming and outgoing problems are 
locally and globally consistent, and \eqref{eq:duality2d} holds.

Note that for concave shapes, some radii of curvature may be negative, which implies 
that the local circle corresponding to that curvature lies outside the cell. In these cases, light rays 
tend to scatter rather than converge. Despite this, a similar analysis can be applied. If the 
test ball is placed at a distance $r$ from the origin and has a small radius $\epsilon$, the near-axis 
approximation holds, and the generalization of Eq.~\eqref{eq:app:A:6} is
\begin{equation}
    \eta = \frac{d}{\epsilon} \approx \left[1 + r \left( 1 - \frac{1}{n} \right)\right]^{-1} < 1.
\end{equation}
This result indicates that the test ball experiences a reduced boost, even before considering light losses.

In the presence of transmission loss, Eq.~\eqref{eq:A9} represents energy conservation and 
applies generally, provided total internal reflection does not trap energy within the cell 
indefinitely. By the duality relation \eqref{eq:duality2d}, Eq.~\eqref{eq:A11} also holds universally: 
if no light path is confined solely by total reflection, the boost for the incoming problem 
remains unchanged. 
However, for structures like a circle, regions near the boundary experience total reflection, leading to a 
reduced average boost there. Likewise for the incoming problem, if the test ball is 
too close to the boundary, \eqref{eq:eta} 
leading to the reduced boost seen in 
Fig.~\ref{fig5}(b). This issue is discussed further at the end of Sc. IV C.

\textbf{F. Averaging for 2D Shapes: A Surface Length Law.} The baseline analysis of a circular 
test ball can be 
extended to arbitrary convex shapes, assuming isotropy in orientation\textemdash that is, the shape samples 
all orientations with equal likelihood. This condition is physically plausible for organisms 
such as algal cells undergoing slow tumbling or Brownian motion. Many unicellular organisms such as 
\emph{Chlamydomonas} and \emph{Volvox} \cite{Goldstein2015} exhibit such dynamics due to flagellar 
motion or ambient fluid fluctuations. These 
lead to statistical averaging over orientations 
on timescales relevant for light exposure.

We parameterize such a shape as $r(\theta)$, with unit outward normal $\textbf{n}(\theta)$, and consider 
illumination from a fixed direction $\textbf{k}_2$. The projected area is given by:
\begin{equation}\label{eq:def:project}
    P = \frac{1}{2}\int_0^{2\pi} d\theta \, \sqrt{g}\, \vert\hat{\textbf{n}}(\theta) \cdot \textbf{k}_2\vert,
\end{equation}
where $g=r^2(\theta)+(dr/d\theta)^2$ is the metric factor, and the prefactor of $1/2$ reflects the 
fact that only the illuminated half contributes to the projection.

To compute the average projected area over all orientations, we rotate the shape by 
$\theta_0$ and average over $\theta_0 \in [0,2\pi]$. Under such a rotation, the shape becomes
\begin{equation}\label{eq:new:shape}
    r_{\theta_0}(\theta) = r(\theta + \theta_0),\quad g_{\theta_0}(\theta) = g(\theta + \theta_0),
\end{equation}
and the normal $\textbf{n}_{\theta_0}(\theta)$ is $\textbf{n}(\theta)$ rotated by $\theta_0$. Letting 
$\textbf{n}'=\textbf{n}(\theta')$, the averaged projection becomes
\begin{align}\label{eq:averaged:projection:2D}
    \bar{P}
    &= \frac{1}{4\pi}\int_0^{2\pi}\!\!\! d\theta' \sqrt{g(\theta')} \int_0^{2\pi}\!\!\! 
    d\theta_0 |\cos(\theta_0+\cos^{-1}{|\textbf{n}' \cdot \textbf{k}_2|})| \nonumber \\
    &= \frac{1}{\pi} \int_0^{2\pi}\!\!\! d\theta \, \sqrt{g(\theta)} = \frac{1}{\pi}L,
\end{align}
where $L$ is the perimeter. Notably, the averaged projected area scales linearly with 
perimeter and is independent of shape geometry. This reinforces that projection-based 
absorption for fluctuating 2D convex bodies respects a perimeter-based law.

\section{Analytical Results for 3D Bodies}
\label{sec:3Danalytical}
The simplest shape 3D shape to analyze is the sphere. 
For the reference ball at the center, 
any incident light ray can be analyzed the same way as in 2D. The boost without loss of light due to 
reflection is analogous to Eq.~\eqref{eq:eta},
\begin{equation}
\eta = \frac{\pi (Rn)^2}{\pi R^2} = n^2.
\end{equation}
Considering loss, Eq.~\eqref{eq:fresnel:eta} becomes
\begin{equation}
\eta = \frac{1}{\pi} \int_0^{\sin^{-1}\delta}d\theta \int_0^{2\pi} d\phi \sin{\theta} \cos{\theta}f(\theta).
\end{equation}

\textbf{B. Averaging in 3D: A Surface Area Law.} The same principle discussed in \S~\ref{sec:2Danalytical}.E 
extends to three-dimensional convex bodies 
with fluctuating orientations. Let the body 
have a fixed surface area and be 
parameterized by its radial function 
$r(\Omega)$, where $\Omega$ denotes a 
direction on the unit sphere, and 
$\textbf{n}(\Omega)$ is the outward unit 
normal. For a fixed direction 
$\textbf{k}_2$ of incoming light, the 
projected area is 
\begin{equation}
    P = \frac{1}{2}\int\! d\theta d\phi \, \sqrt{g}(\theta,\phi) \, \vert\textbf{n}(\theta,\phi) 
    \cdot \textbf{k}_2\vert,
\end{equation}
where the factor $1/2$ counts
the illuminated half of the body, and the
metric factor is
$\sqrt{g}(\theta,\phi)=r\sqrt{[r^2+(dr/d\theta)^2]
\sin^2\theta+(dr/d\phi)^2}$.

To compute the orientational average 
of $P$, we integral over 
all rotations $\Omega_0$. 
Without loss of generality, take $\mathbf{k}_2$ to align with the polar axis, and replicate the steps leading to
\eqref{eq:averaged:projection:2D}, with 
$d\Omega_0=d\phi_0d\theta_0\sin{\theta_0}$,
yielding
\begin{align}
    \nonumber\bar{P} &= \frac{1}{2} \int \frac{d\Omega_0}{4\pi} \int \frac{d\Omega}{4\pi} \, \sqrt{g}_{\Omega_0}(\Omega) \, |\textbf{n}_{\Omega_0}(\Omega) \cdot \textbf{k}_2| \\\nonumber
    &=\frac{1}{2}\int d\Omega' \, \sqrt{g}(\Omega')\int_0^{2\pi}\frac{d\phi_0}{2\pi} \int_0^\pi \frac{d\theta_0 \sin\theta_0}{2} \\ 
    &\quad\quad  \times \vert\cos(\theta_0+\cos^{-1}(\textbf{n}(\Omega) \cdot \textbf{k}_2)\vert\nonumber \\
    &= \frac{1}{4}A,
\end{align}
where $A$ is the object's surface area. 
As in the 2D case, this result 
depends only on total 
surface area. The average projected area of 
a convex 3D object under uniform 
orientation fluctuations thus follows a 
surface area law.

Also as in 2D, the above analysis breaks down for concave shapes. In those cases, parts of the 
surface can shade each other, leading to a reduction in the effective projected area. 
The impact of self-shadowing must be accounted for separately, as it can significantly 
alter the absorption characteristics and invalidate the simple surface-area scaling seen 
in convex geometries. We analyze such effects 
numerically in \S~\ref{sec:3Dnumerical}.

{\textbf{C. {\'E}tendue and Flux Conservation.}
{\'E}tendue (often denoted $\mathcal{E}$) can be viewed as the ``phase space volume" 
of a light beam \cite{Chaves}. 
It quantifies the spread of light in both position and direction. For a beam passing through a 
cross-sectional area $A$ and contained within a solid angle $\Omega$, the {\'e}tendue is 
$E = A \Omega$.
In media where the refractive index varies spatially, this generalizes to
\begin{equation}
    \mathcal{E} = n^2 A \Omega,
\end{equation}
incorporating refractive effects on the direction of light rays. $\mathcal{E}$ is conserved 
in passive optical systems, as 
we now show follows from flux considerations.

Assume a beam of light crosses the interface without loss and consider a corresponding 
set of rays defined in 
medium $1$ by an area element $dA_1$, with chief ray direction $\Omega_1$ and solid angle 
element $d\Omega_1$. 
After refraction into medium $2$, these same rays will pass through some (generally different) area element 
$dA_2$ and subtend a solid angle $d\Omega_2$ with direction $\Omega_2$. To connect 
$d\Omega_1$ and $d\Omega_2$, 
we must understand how a cone of rays in medium 1 maps into medium 2.  From the differential 
of solid angle in spherical
coordinates,
\begin{equation}
d\Omega = \sin\theta \, d\theta \, d\phi,
\end{equation}
we differentiate Snell's law and obtain 
\begin{equation}
    n\cos\theta_t d\theta_t=\cos\theta_i d\theta_i.
\end{equation}
Since reflection does not change the azimuthal angle, we can multiply by $d\phi$ on both 
sides and use Snell's law again
to obtain
\begin{equation}
\cos\theta_td\Omega_2 = \frac{1}{n^2}\cos\theta_i d\Omega_1.
\end{equation}
The light rays intersect the boundary in an area element $dA$, and if we define $dA_1^\perp=dA\cos\theta_i$ 
and $dA_2^\perp=dA\cos\theta_t$, we find
\begin{equation}
    \frac{dA_1^\perp}{dA_2^\perp}=n^2\frac{d\Omega_2}{d\Omega_1}.
    \label{dA}
\end{equation}
Under the chief ray approximation, 
each point in the area element has the same solid angle. The LHS of \eqref{dA} is recognized as the boost factor 
for the incoming problem and the RHS as that of the outgoing problem. We thus establish that for each solid angle 
direction $\Omega_1$ for the outgoing problem,}
\begin{equation}\label{eq:duality}
    n^2\eta_B(\Omega_1)=\eta_P(\Omega_2).
\end{equation}
where $\eta_B$ and $\eta_P$ are boosts in the bioluminescence (outgoing) and photosynthesis (incoming) cases defined in 
\eqref{eq:app:13} and~\eqref{eq:app:19}. The duality in\eqref{eq:duality} is numerically verified in 
the next section, where the outgoing and incoming configurations yield equivalent intensity distributions.

In the same spirit as Eq.~\eqref{eq:A11}, we may calculate the average boost for the photosynthesis problem 
by integrating over incoming angles, 
\begin{align}
\eta_P^{\text{avg}}&=\frac{1}{4\pi}\int d\Omega_2\eta_P(\Omega_2)=\frac{1}{4\pi}\int d\Omega_1 \frac{\eta_P(\Omega_2(\Omega_1))}{\eta_B(\Omega_1)}\nonumber \\&=\frac{1}{4\pi}\int d\Omega_1 n^2=n^2,
\label{eq:eta:avg 3D}
\end{align}
where in the second line we have used $\eta_B(\Omega_1)=\partial \Omega_2/\partial \Omega_1$ as the Jacobian. 
The average boost is only a function of the relative reflective index $n$, and is irrespective of the 
concave shell shape and the location of the test ball. 

In fact, the argument above works for the case of multiple refractions, and when accounting for transmission losses as in 
Eq.~\eqref{eq:fresnel}. For the incoming problem, a principal light ray hits the test ball after the initial refraction 
and further reflections; we can trace back the ray from the test ball outside, which becomes the outgoing problem. 
The conservation of {\'e}tendue ensures that the duality of the boost hold at any point. The loss at the 
interface is also the same. 

In considering all reflections and loss, let us label the boost by the number of reflections $i$ that have 
occurred thus far. The outgoing boost is 
$\eta_B^i(\Omega_1)=f^i(\Omega_1)d\Omega_2/d\Omega_1$, where $f_i(\Omega_1)=\Pi_{m=1}^i(1-f(\theta_r^m))f(\theta_t)$ 
accounts for the loss according to Eq.~\eqref{eq:fresnel}. 
We similarly modify the definition of the incoming boost to account for multiple reflections, including losses. 
For the outgoing problem, all light will exit the body after a sufficiently large number of reflections, unless 
the situation is highly symmetric and some 
light rays are trapped by total internal reflection (See Fig.~\ref{fig9}(b) where, near the edge, the average boost reduces to below 1).
Then, energy conservation gives $\sum_i f^i(\Omega)=1$, and thus
\begin{align}
    \eta_B^{\text{avg}}\!&=\!\int\! \frac{d\Omega_1}{4\pi}\sum_{i}\! \eta_B^i(\Omega_1)
    \! =\! \int \frac{d\Omega_2}{4\pi} \sum_i f^i(\Omega_1)=1.
    \end{align}
Since losses for incoming and outgoing light rays are the same in each direction, 
the \eqref{eq:duality} now reads $n^2\eta^i_B(\Omega_1)=\eta_P^i(\Omega_2^i(\Omega_1))$. 
Denote the direction of the outgoing light ray after $i$ reflections as $\Omega_2^i(\Omega_1)$. 
Eq.~\eqref{eq:eta:avg 3D} becomes
\begin{align}\label{eq:eta_3D ref}
    \eta_P^{\text{avg}}
    &=\sum_i\int \frac{d\Omega^i_2}{4\pi} \eta^i_P(\Omega^i_2)
    =\frac{1}{4\pi}\sum_i\int d\Omega_2^i n^2\eta_B^i(\Omega_1) \nonumber \\
     &=\sum_i\frac{1}{4\pi}\int d\Omega_1 n^2f^i(\Omega_1)=n^2.
\end{align}

\section{Numerical Results for 3D Bodies}
\label{sec:3Dnumerical}

\begin{figure*}[t]
    \includegraphics[width=0.90\linewidth]{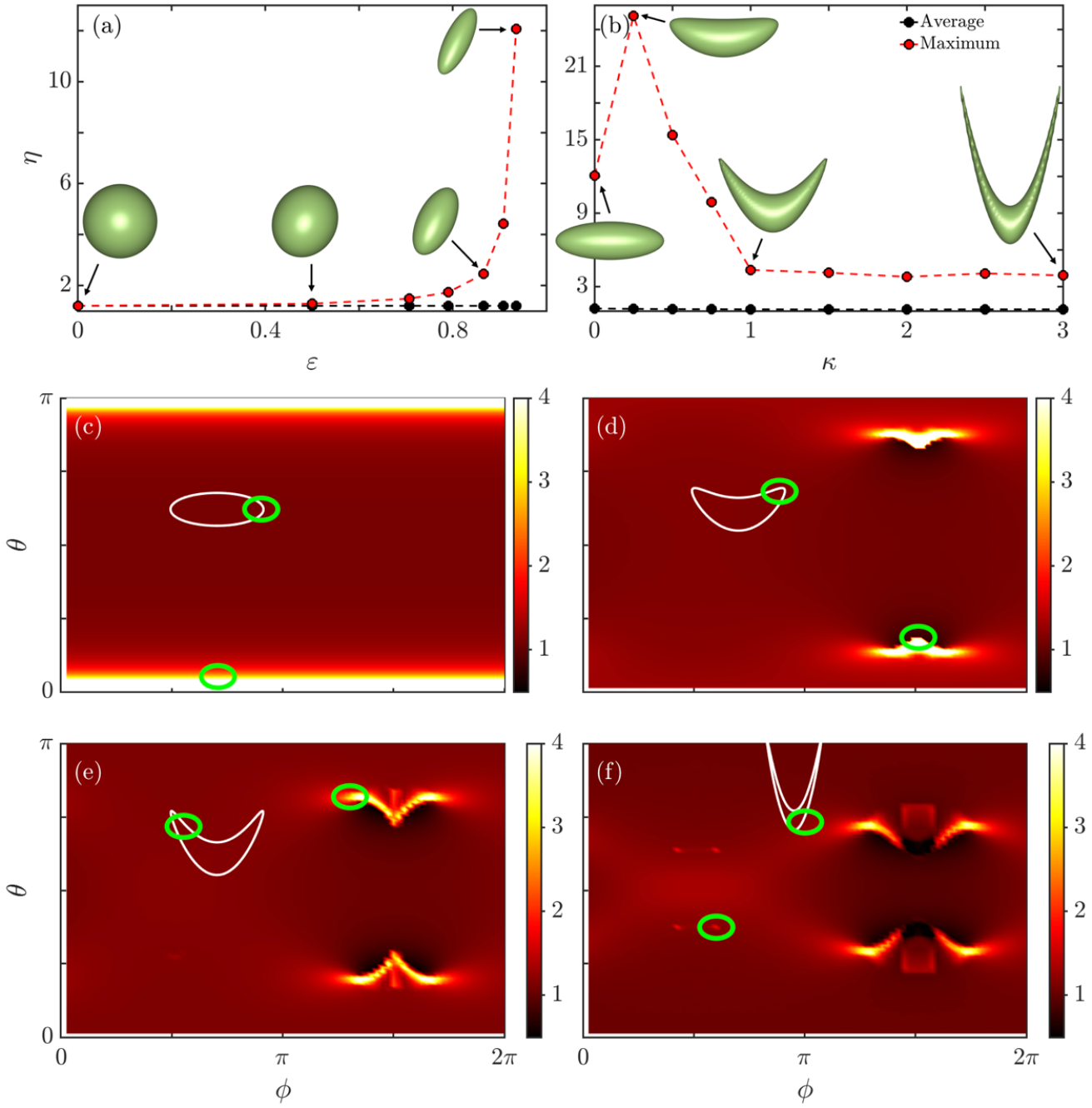}
    \caption{Variation with cell shape of maximum intensity boost at the center of the body. 
    (a) Light intensity observed at the 
    center of ellipses increases as a function of eccentricity $\varepsilon$, while the average intensity 
    remains constant. (b) As a function of the curvature parameter $\kappa$ the 
    maximum boost for bent ellipsoids is 
    nonmonotonic.  (c-f) Intensity boost as a function of angles $(\theta,\phi)$. For an ellipsoid, 
    axisymmetry 
    results in a boost that is independent of $\phi$. Parameter values are $\kappa=0,0.5, 1, 3$ 
    and eccentricity is $\varepsilon=\sqrt{7/8}$. For bent shapes (d-f), the maximum stays near the 
    direction of the tips. The green circles in (c-e) indicates the entry region giving strong boost. Note the focusing of the first reflections circled in green in (f).}
    \label{fig7}
\end{figure*}

In this section, we present numerical results that support the conclusions drawn in the analytical sections above. 
We consider a two-parameter family of shapes for numerical computations.  Built around the parameterization of
ellipsoids, these shapes interpolate 
between the sphere (a suitable model for \textit{Chlamydomonas}), and those with eccentricity approaching
unity (appropriate to \textit{P. fusiformis}) and then proceed to ellipsoids bent around their 
major axis (a shape like \textit{P. lunula}).  
In a system of units made dimensionless by the semi-major axis $a$ of the ellipsoidal limit, 
this family can be written as
\begin{equation}\label{eq:shape:lunula}
    x^2+ \frac{1}{1-\varepsilon^2}\left[y^2 + \left(z-\kappa x^2\right)^2\right]= 1,
\end{equation}
where $\varepsilon=\sqrt{1-(b/a)^2}$ is the eccentricity and $b$ is the semi-minor axis of the 
limiting ellipsoids obtained when $\kappa=0$.  These have major axes lying
in the $xz$-plane with a radius $r(\theta,\phi)$ measured from the ellipsoid center of
\begin{equation}
    r(\theta,\phi)=\frac{\sqrt{1-\varepsilon^2}}{\sqrt{1-\varepsilon^2\cos^2\theta}},
\end{equation}
while for $\kappa\ne 0$ the shapes are bent in the $xz$-plane around the curve 
$x=\kappa x^2$ for $x\in [-1,1]$,
\begin{equation}
    z=\kappa x^2 \pm \sqrt{1-\varepsilon^2}\sqrt{1-x^2}.
\end{equation}
Figure~\ref{fig7} shows such shapes for various values of $\varepsilon$ and $\kappa$.

\textbf{A. Photosynthesis at the Center.} To examine how cell shape influences light concentration near the 
center, we model the chloroplast as a small, perfectly absorbing sphere at the center of the cell, shown
as the green circle in Fig.~\ref{fig3}(a). Incoming parallel light rays are assumed to arrive uniformly from 
all directions. 
We discretize angular directions and 
emit rays from a plane perpendicular to each specified propagation direction $\mathbf{k}_1$. Here and
in following sections geometric symmetries allow us to 
restrict the angular sampling to half of the spherical 
polar angle domains in $\phi$ and $\theta$. 
When the cell is bent, only two planar symmetries remain, resulting in boost-angle profiles with mirror symmetry 
across $\phi = \pi/2$ (and $3\pi/2$) and $\theta = \pi/2$.
Both $\phi$ 
and $\theta$ are discretized into $40$ sampling points. 
For each direction, $360,000$ rays are generated so 
that the spacing between rays is smaller than the 
radius of the test ball, while ensuring that the ball remains small relative to the cell geometry.

Computational constraints restrict the region of ray 
entry to a zone centered on the test ball with a width 
$15$ times its radius\textemdash an approximation that
remains accurate, as confirmed by the close match between computed intensities in Fig.~\ref{fig7}(a,b) and the analytical prediction \eqref{eq:eta_3D ref}.
Each ray carries equal energy, undergoes refraction at the cell boundary according to Snell’s law \eqref{def:snell}, and is attenuated by its Fresnel transmission coefficient \eqref{eq:fresnel}. 
Each ray undergoes up to 11 reflections and 
those whose energy drops below $10\%$ 
of their initial value are discarded to reduce computation time. This setup leads to billions of ray computations for each test ball position.
The total absorbed energy is computed and compared to a case without a surrounding cell boundary.

We focus on two metrics: the maximum intensity boost (along the optimal direction) and the 
average boost (over all directions). In the spherical case, symmetry ensures that $\eta$ 
is direction-independent, so the average and maximum coincide. As the eccentricity $\varepsilon$ 
increases, Fig.~\ref{fig7}(a) shows that a preferred direction emerges, with the 
maximum intensity increasing dramatically, from $\sim\!1.2$ to $\sim\!12$. Introducing 
bending with a finite $\kappa$ (Fig. \ref{fig7}(b)) further increases the peak, which exceeds 
$25$ before declining to $4$ . This non-monotonic trend arises from the interplay between the 
increasing distance between the tip and the center, and decreasing radius of curvature radius at the tip. 
While the initial bending aligns focal zones with the center (by slightly increasing the axis 
length as initially the center lies between the focus and the tip), enhancing light concentration,
further bending misaligns them, decreasing the central boost.

The average intensity is less sensitive to shape. For the sphere and ellipsoids, it remains near 
$n^2\simeq 1.2$ as predicted by \eqref{eq:eta_3D ref}. Bending decreases the average slightly, especially 
near the concave parts of the 
shell, due to redirection of light rays away from the center as $\kappa$ increases.

\textbf{B. Angular Profile of the Boost.}
To understand better the entry point of the maximum boost and the distribution of light intensity, 
the boost observed at the center of the body is analyzed as a function of the incoming angles ($\theta,\phi$). 
For ellipsoids, the boost profile in Fig.~\ref{fig7}(c) exhibits the expected axisymmetry in $\phi$. 

For mild bending, Fig.~\ref{fig7}(d) shows that the peak region is centered at the tip of 
the bent shape and is the brightest spot in the profile. As bending increases in panels (e) and (f), 
light from the tip no longer focuses directly on the center, but rays entering from nearby directions 
do. This shift in the angular location of intensity peaks is accompanied by a migration of the 
$\theta$-center of the bright regions, as the tip curvature becomes sharper.

As the shape becomes more extreme (panel f), additional bright spots and faint striping patterns emerge 
due to internal reflections of rays within the curved shell. The bright spots correspond to 
rays that enter the cell, reflect internally at the tip, and are redirected toward the center. The faint 
stripes are remnants of these multiply reflected paths. These reflection-induced features become 
especially prominent when the tip region is sharply curved and capable of directing incident rays back inward.

Dark regions in the angular map correspond to incident directions that either miss the central test 
ball or undergo reflection away from it. These regions typically have a boost near 1, indicating 
that the light reaching the test ball is neither enhanced nor significantly diminished.

\begin{figure}[t]
    \includegraphics[width=\columnwidth]{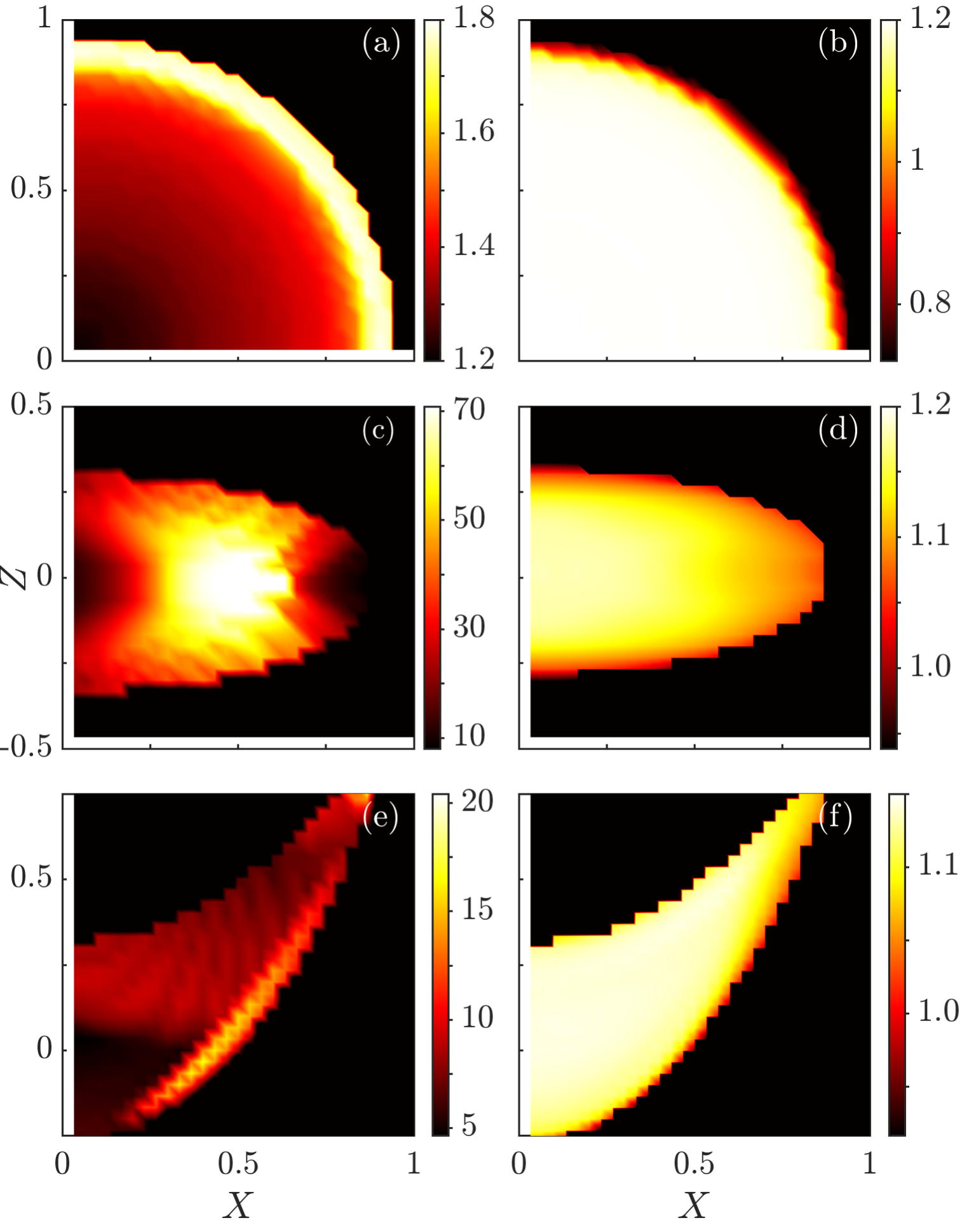}
    \caption{Spatial distribution of intensity boost. 
    (a,b) The sphere exhibits a spherically symmetric profile. (c,d) Ellipsoid with $\varepsilon=\sqrt{7/8}$. 
    (e,f) \textit{P. lunula} shape with $\kappa=1.$} 
\label{fig8}
\end{figure}

\textbf{C. Spatial Distribution of the Boost.}
To gain further insight into the light intensity profile\textemdash and to explore the potential 
biological advantages of non-axisymmetric shapes and organelle distributions\textemdash 
we examine the spatial variation of light intensity within the cell. Specifically, we analyze both the 
average and the maximum intensity boost at various locations. The results are presented in Fig.~\ref{fig8}.
All plots consider only a smaller region inside the cell, as edge effects near the boundary make calculations 
unreliable. 

For the sphere, $40$ sample points are taken along the 
radial direction, with the intensity distribution being invariant in the azimuthal angle $\phi$. Therefore, discretizing the polar angle $\theta$ into $40$ intervals is sufficient.
For the ellipsoid, $208$ sampling points are distributed 
inside the cell, leveraging symmetry along the $X$-axis. 
However, due to the reduced symmetry of this geometry, 
angular profiles are only symmetric about $\phi = \pi/2$. 
The high curvature of the ellipsoid surface demands a 
significantly finer angular grid; coarse sampling results in
stripe-like artifacts—an issue commonly encountered in ray 
tracing simulations \cite{Modest2013}. To eliminate these artifacts in the computed maximum intensity distribution, we
discretize the angular space using $160$ points in $\theta$ and $80$ in $\phi$. We use $8,100$ rays per direction, which has been found to yield accurate results in practice.
For the \textit{lunula} example, $387$ sample points 
are used inside the shape, employing the same angular 
discretization and simulation parameters as with 
the ellipsoid. However, due to the fourth-order nature of 
the parametrizing equation~\eqref{eq:shape:lunula}, the simulation requires substantially more computational time.

\begin{figure}[t]
    \includegraphics[width=0.98\columnwidth]{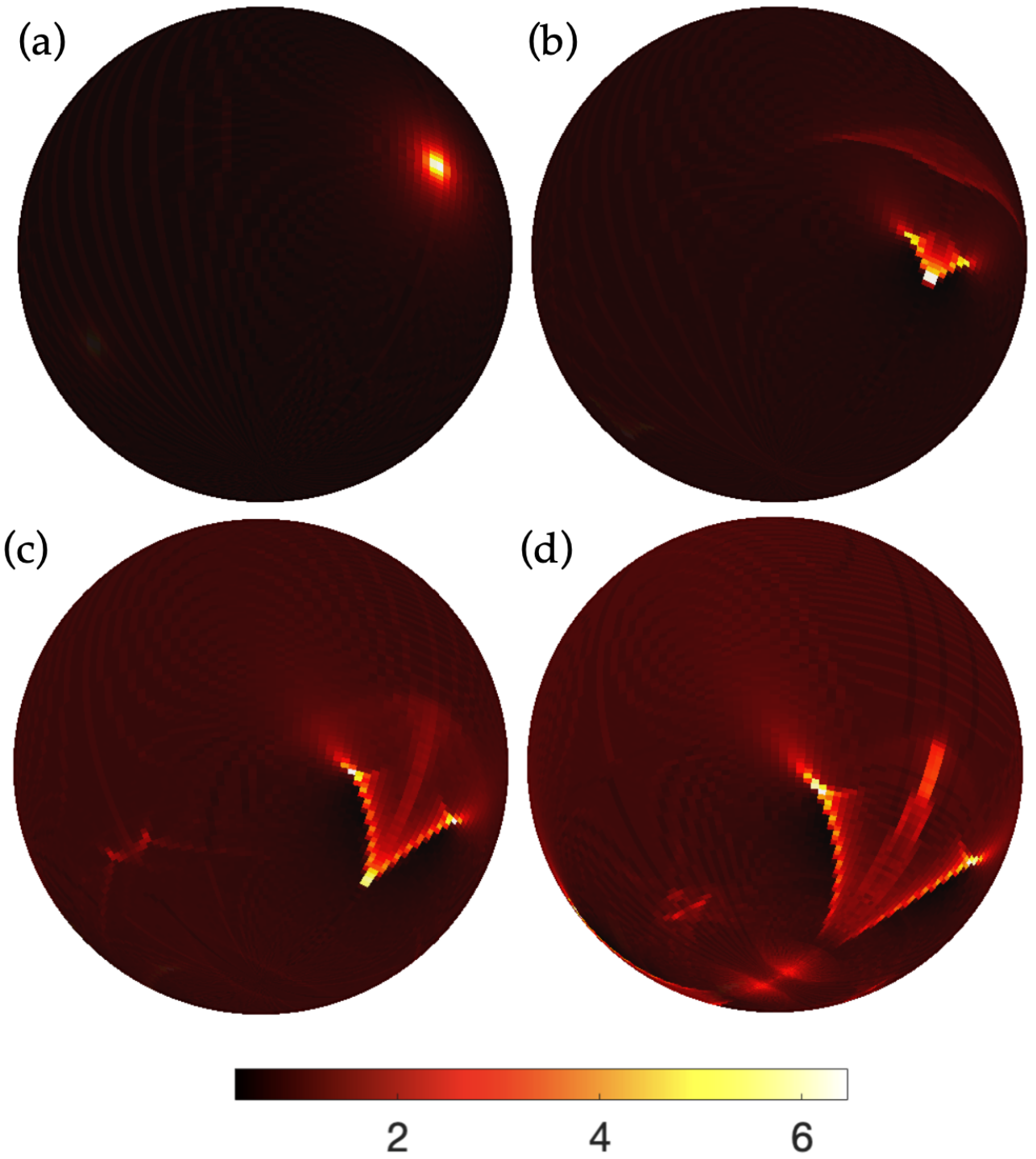}
    \caption{Angular distribution of light emission 
    from the center of a cell. For a source fixed at the center and with eccentricity $\varepsilon=\sqrt{7/8}$. The bending 
    parameter $\kappa=0, 0.5, 1$ and $2$ in (a-d). The boost region enlarges with $\kappa$ while 
    the maximum intensity first increases then decreases, a trend also seen in Fig.~\ref{fig7}(b). The patterns resemble those of Figs.~\ref{fig7}(c-f).}
\label{fig9}
\end{figure}

Within the sphere, the averaged intensity is highly uniform, with a boost of $\sim 1.2$ as discussed at the end of Sec. IV C. and
as predicted by Eq.~\eqref{eq:eta_3D ref}, until it drops to $<0.8$ near the boundary. 
Variations are likely due to numerical errors and 
the fact that only finitely many reflected rays are considered, supporting the conjecture that 
all points within a convex shape experience the same average intensity. The maximum intensity 
is lowest at the center, increasing to $\sim 1.7$ near the boundary. As the test ball approaches 
the boundary, the density profile becomes more uneven due to the proximity to the focal region 
of light from the antipodal point, as well as the increasing effectiveness of internal reflections. 
This is larger than the value calculated in Eq.~\eqref{3Dboost} due to consideration of internal reflections.

A similar pattern is observed for the ellipsoidal case. Points near the focal region experience 
a maximum intensity boost exceeding $70$, as light accumulates significantly near the focal point. 
The averaged intensity remains constant near the center, 
decreasing slightly from $\sim 1.18$ (close to the expected value of $n^2$) by $\sim 4\%$, then quickly drops to below $1$, again likely due to the 
limited number of rays considered.
For the \textit{lunula} shape, the maximum intensity also shows sensitivity to discretization. However, 
the maximum boost is generally weaker than in the ellipsoidal case. The thin bright line 
in Fig.~\ref{fig9}(e) results from a focusing effect at the tip. The average intensity remains nearly 
uniform near the center, decreasing from $\sim1.14$ to $\sim1.1$ near the tip due to the concavity of the shape. We also observe a decrease to below $1$ near the boundary, similar to the ellipsoid case, likely due to limited light ray regions considered and the energy cutoff of light rays.

Overall, while the average intensity is robust across shapes, the spatial distribution of maxima 
reveals how cell geometry can create strong local variation. These focal hot spots highlight how even simple geometric differences can significantly modulate internal light fields. 

\textbf{D. Bioluminescence.} To model the outgoing problem, a point source emits light uniformly 
from a location inside a given cell shape. Light rays are refracted outward and reflected internally 
within the cell. Since the cell is small compared to the inter-organism distance, the location of 
outgoing rays can be ignored, and the light intensity distribution is evaluated on an infinite sphere 
centered at the cell, plotted with respect to the solid angle. The calculation is normalized 
against the case of a point light source at the center of the observation sphere. 
The computational process is simplified by generating light rays solely from the test sphere. The spherical grid is discretized into $160 \times 160$ points in $\theta$ and $\phi$, respectively, and $409,600$ test light rays are generated from the ball. For reflections, we disregard those carrying less than 1\% of their initial energy.
The light intensity distribution at each location only requires millions of light ray realizations and
achieves better angular resolution. Future studies on the spatial distribution of photosynthetic 
boost may benefit from employing bioluminescence-based simulation techniques.
The light profile 
where the test body is placed at the center of various cell shapes is produced in Fig.~\ref{fig9}, 
which shares the same pattern as Figs. \ref{fig7}(c-f). This further confirms the duality in 
Eq.~\eqref{eq:duality}.

\begin{figure}[t]
    \centering
\includegraphics[trim={40 530 50 60}, clip,width=0.98\columnwidth]{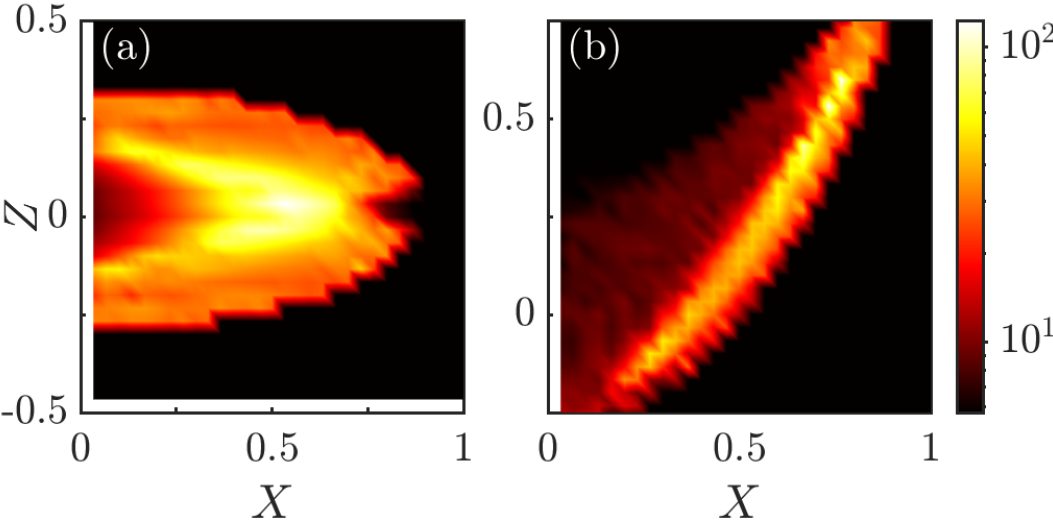}
    \caption{Spatial distribution of maximum focus boost. (a,b) correspond to (c,e) in Fig. \ref{fig8}. 
    The plot is in log scale, with bright yellow regions showing strong focusing.}
    \label{fig:10}
\end{figure}

We similarly examine the spatial distribution of the maximum intensity, as shown in Fig.~\ref{fig:10}. 
The maximum intensity values at each point are higher since, at the same angular resolution, the outgoing 
simulation converges more rapidly to the maximum. 
Notably, the bright regions remain consistent, in 
agreement with Eq.~\eqref{eq:duality}. 

It is worth noting that our simulations of the 
incoming and outgoing problems handle concave shapes 
slightly differently. Specifically, light rays that
re-enter the cell after reflection/refraction are not included 
in the analysis. These rays are expected to have 
negligible impact due to significant energy loss, 
as described by Eq.~\eqref{eq:fresnel}. Nevertheless, a more refined numerical treatment could be applied for concave geometries such as that of \textit{P. lunula}.

\section{Discussion}
\label{sec:discussion}

We have presented a framework to understand how cell shape influences the distribution 
of light in organisms. Our work covers both incoming light, relevant for photosynthesis, 
and outgoing light, relevant for bioluminescence, and is based on geometrical optics. We introduced the 
notion of a boost factor $\eta$ to quantify the focusing or defocusing of light within a cell. For simple geometries like the circle and the sphere, we obtained exact analytical results. In 
more complex shapes such as ellipses and bent geometries motivated by real dinoflagellates, we computed 
the intensity distributions numerically. In all cases, we find that geometric focusing alone can produce 
significant spatial variation in the light field, even in the absence of any specialized optical structures.

A key observation is that incoming and outgoing problems are related by a geometric duality, derived 
from conservation of {\'e}tendue in a passive optical system~\cite{Chaves}. Duality allows us to deduce 
the intensity distribution in one case from that in the other. The duality relation was 
verified numerically across a range of geometries, and we showed that both configurations 
yield equivalent angular profiles after appropriate transformation.

Our results show that transparent, weakly refracting cells can experience substantial spatial 
variation in internal light intensity due to geometry alone. In bent cells, the light can be 
concentrated near the outer curvature by more than an order of magnitude. For convex geometries, 
the average boost is close to that of the circle or sphere, indicating that geometry tends to 
redistribute light rather than produce an amplification or reduction.

These results suggest that cells can passively 
manipulate light intensity using shape alone and may 
be relevant to how cells sense or respond to light. 
For photosynthetic cells, strategic localization of 
chloroplasts near focal zones (e.g., tips or curved 
regions) could maximize light absorption 
~\cite{bjorn2008photobiology, wada2003chloroplast}. 
For bioluminescent organisms, shaping the cell to 
direct emission could enhance signal directionality. 
For instance, our ray tracing simulations show that 
\textit{P. fusiformis} emits light more strongly 
in the lateral directions than axially (Fig.\ref{fig2}), 
perhaps confining in-plane signaling among neighboring 
organisms while reducing visibility from predators 
above or below, while \textit{P.~lunula} exhibits a 
concave-side bias and axial focusing, particularly when 
the source is centrally located. These anisotropies 
could influence signal projection, enhancing visibility 
in preferred directions while reducing the detection 
risk in others. Combining these directionalities with 
the inherent tumbling dynamics of elongated objects in 
shear flows, we may view cells as ``stochastic beacons"
of light, leading to nontrivial fluctuation statistics 
of light production in large groups of organisms.
Recent studies of dinoflagellate light production and 
its cellular mechanisms suggest that geometric effects 
may also influence the triggering process leading to 
flashes~\cite{Jalaal}. Tools such as high-speed imaging 
or optogenetic reporters may provide ways to probe the 
internal light distribution in 
vivo~\cite{airan2009temporally}. Our findings may 
help explain why certain phytoplankton 
have evolved strongly eccentric or bent geometries, 
and they offer a testable hypothesis for 
subcellular organization driven by optical advantages. 

Our model assumes geometrical optics and 
a homogeneous refractive index inside the cell. 
These assumptions are valid for cells much larger than the wavelength of light, but effects such as 
diffraction or interference may become relevant for smaller cells or near high-curvature regions. The 
presence of strongly absorbing components like pigments or organelles could also modify the results;
the inclusion of absorption is an important future extension. 
Likewise, we have not modeled time-dependent effects such as cell rotation or deformation, which may alter 
the effective sampling of the light field.
We also restricted attention to passive optical structures, neglecting any active light-guiding mechanisms 
or structural features. In some species, there is evidence of microlenses, reflective layers, or cellular-scale waveguides~\cite{johnsen2012optics}, which could be treated 
by extending the present framework to include multiple layers or graded refractive index profiles.

Finally, it is possible to couple the boost factor to light-activated processes, such 
as chloroplast migration or flash triggering. And 
the duality between incoming and 
outgoing problems may have implications for how phototactic and bioluminescent behaviors are coordinated 
in organisms that rely on both.

\begin{acknowledgments}
We are grateful to M.V. Berry for discussions at an early stage of this research, and to 
Mazi Jalaal for the image in Fig. \ref{fig1}(b).  This work was supported 
in part by a Trinity College Summer Studentship and 
a Gates Scholarship (MY) and
Grant No. 7523 from the Gordon and Betty Moore Foundation (SKB \& REG).
\end{acknowledgments}

The data that support the findings of this article are openly available \cite{Zenodo}.

\end{document}